\documentclass[prl,twocolumn,floatfix,superscriptaddress]{revtex4-1}
\usepackage{dcolumn,amsmath}
\usepackage{graphicx}
\usepackage{bm}
\usepackage{hyperref}

\newcommand{\Eeff}{\ensuremath{E_{\rm eff}}}
\newcommand{\eEDM}{{\em e}EDM}

\begin{document}
   \title{Reply to the Comment on ``Theoretical study of thorium monoxide for the electron electric dipole moment search: Electronic properties of $H^3\Delta_1$ in ThO''}

\author{L.V.\ Skripnikov}\email{leonidos239@gmail.com}
\author{A.V.\ Titov}
\homepage{http://www.qchem.pnpi.spb.ru}
  \affiliation{B.P.~Konstantinov Petersburg Nuclear Physics Institute, Gatchina, Leningrad district 188300, Russia}
\affiliation{Dept.\ of Physics, Saint Petersburg State University, Saint Petersburg, Petrodvoretz 198504, Russia}
\date{10.10.2016}

\begin{abstract}
We reply to the comment [M. Denis, T. Fleig, arXiv:1605.03091v1 (2016)] on paper [L.V. Skripnikov and A.V. Titov, J. Chem. Phys. 142, 024301 (2015)].
\end{abstract}

\maketitle

\section{Introduction.}

Recently a new progress in the search for the electron electric dipole moment (\eEDM) has been obtained in the ThO beam experiment \cite{ACME:14a}. Originally, the interpretation of the experiment \cite{ACME:14a} in terms of \eEDM\ has been based on the theoretical data from \cite{Skripnikov:13c}, where the parameter called effective electric field (\Eeff) has been calculated with 15\% of theoretical uncertainty. In the study the electron correlation was treated within the coupled cluster approach with explicit consideration of the 38 outer-core and valence electrons of the ThO molecule. After that a new study of the ThO electronic structure was published by T. Fleig and M.K. Nayak in Ref. \cite{Fleig:14}. There the final 
value of \Eeff\ was calculated within the multireference configuration interaction approach with explicit treatment of the 18 valence electrons of ThO. The stated theoretical accuracy was estimated to be 3\%.
In that paper a comparison of different methods to treat electron correlation in the case of ThO system have been given, e.g. comparison of the multireference configuration interaction with single and double excitations (MR-CISD) method with small active space (3 Kramers pairs in active space) with the single-reference coupled cluster with single and double cluster amplitudes (CCSD).
Inspired by the paper \cite{Fleig:14} we have performed a new study of ThO electronic structure. The main goal was to decrease theoretical uncertainty of our work \cite{Skripnikov:15a} and perform a more detailed comparison of different methods. For this a new series of single-reference and multireference coupled cluster (CC) as well as configuration interaction (CI) calculations has been performed. We have successively estimated contributions to the considered properties from different types of correlation effects. From the obtained data we have also made some estimations of accuracy of the methods used in \cite{Fleig:14} and concluded that the uncertainty of 3\% stated in \cite{Fleig:14} was notably underestimated.
Recently, M. Denis and T. Fleig published comment \cite{FleigComment:16} on the paper \cite{Skripnikov:15a} 
where a criticism made in \cite{Skripnikov:15a} about the methods used in \cite{Fleig:14} was addressed.

Here we give our reply to the comment.

\section{Discussions}

\subsection{Spinor basis set}

The state of interest in ThO for the experiment to search for the \eEDM\ is the lowest lying $H^3\Delta_1$ excited electronic state. Qualitatively, this is the state with two unpaired electrons occupying $7s_\sigma$ and $6d_\delta$ orbitals localized mostly on Th atom, i.e. the $^3\Delta_1$ state of ThO has $7s_\sigma^1 6d_\delta^1$ configuration. The ground state of ThO, $^1\Sigma^+$, has no open shells and the outermost (valence) electrons occupy the $7s_\sigma^2$ shell.
As one of the tests for different methods of treatment of electron correlation we have advanced the saturation (stability) of the results with respect to the different choice of one-electron functions from which one builds many-electron Slater determinants to be used in the wave-function expansion in correlation treatment. The best possible approach (that is the full configuration interaction approach, Full-CI) should be, in fact, completely irrelevant to the choice of the one-electron basis functions while approximate methods such as truncated CI and CC approached will be. The best choice of one-electron functions is complicated and cannot be unambiguously answered. As pointed in \cite{Skripnikov:15a} one can use spinors produced in the Hartree-Fock method, complete active space self-consistent field method, natural spinors produced by diagonalizing the one-electron density matrix obtained at some preliminary (previous iteration, etc.) correlation calculation \cite{Cossel:12}. Besides, one can use Brueckner orbitals, etc.
One of the examples is the quasi-restricted Hartree-Fock approach, where one produces orbitals for a certain electronic state but use them to construct the Fermi vacuum state for other electronic state. There are many points which can be considered when one selects the particular set of one-electron functions, such as physical motivation \cite{FleigComment:16}, convergence motivation, etc. However, the choice is only relevant to approximate methods and irrelevant to exact full-CI method. 
To compare stability of different correlation methods we have generated two sets of orbitals within different versions of the Hartree-Fock method. One of them was produced for the closed-shell ground electronic state of ThO and the second one was produced by the average-of-configuration Hartree-Fock calculation for two electrons in the six spinors (three Kramers pairs). The latter correspond to $7s_\sigma$ and $6d_\delta$ of Th.

In the paper \cite{Fleig:14} the authors used the multireference configuration interaction methods to treat electron correlation. One of the series of calculations denoted as MR$_K$-CISD used active space composed from occupied open-shell orbitals and a number of virtual orbitals of ThO. For example, the active space of the MR$_3$-CISD method included 3 Kramers pairs: $7s_\sigma$ and two $6d_\delta$. The final calculation of \cite{Fleig:14} included  9 additional virtual Kramers pairs resulting in the MR$_{12}$-CISD approach. In \cite{Skripnikov:15a} we have treated the simplest considered MR-CI method, MR$_3$-CISD, and compared it with the single-reference coupled cluster methods to further investigate the original comparison of these methods in \cite{Fleig:14}.
It was found that the uncertainty of \Eeff\ and hyperfine structure (HFS) constant, A$_{||}$, due to the choice of one-electron basis functions for the case of the MR$_3$-CISD method is rather high (see Table II of \cite{Skripnikov:15a}) while for the used CC approaches it is negligible.
The uncertainties of the \Eeff\ and A$_{||}$ values calculated within the MR$_3$-CISD method was found to be  7\% and 11\%, respectively. Basing on the result (as well as others) we chose the single-reference CC methods as our basic methods in \cite{Skripnikov:15a}.
One should note that we did not state that the uncertainly of MR$_K$-CISD with $K>3$ will be the same as in the MR$_3$-CISD case. Obviously, the former methods accounts electron correlation better than the latter, therefore in general they should be closer to the full-CI case which is irrelevant to the choice of one-electron basis set. This was confirmed in \cite{FleigComment:16}.
According to Table I of \cite{FleigComment:16} the effective electric field is almost irrelevant to the choice of one-electron basis functions for the case of the MR$_{12}$-CISD method.
However, the uncertainty of a natural test parameter, HFS constant, is still has some uncertainty of 7\% which does smaller than in the case of MR$_3$-CISD mentioned above. 

\subsection{Basis set}

In \cite{FleigComment:16} the authors state: ``${P,T}$-odd constants are also insensitive to basis set enlargement within the 4c-MR$_{12}$-CISD(18) model, in contrast to what has been asserted by Skripnikov and Titov''. Actually, it was only stated in \cite{Skripnikov:15a} that estimation of uncertainty on the basis set performed in \cite{Fleig:14} was based on the MR$_{3}$-CISD approach and found to be small. On the basis of data in \cite{Fleig:14} we noted \cite{Skripnikov:15a} that within a more accurate MR$_9$-CISD(18) method the values of \Eeff\ differ by about 2.9 GV/cm for the vTZ vs.\ vDZ basis sets, which is not negligible.
In the new series of calculations performed by M. Denis and T. Fleig  presented in Table I of \cite{FleigComment:16} one can see that the dependence on the choice of the basis set of the MR$_{12}$-CISD(18) method is smaller than for the MR$_9$-CISD(18) published in \cite{Fleig:14}.
Indeed the difference of the MR$_{12}$-CISD(18) values of \Eeff\ within vDZ and vTZ basis sets is only 0.1 GV/cm.
Thus, basing only on the data from Refs. \cite{FleigComment:16,Fleig:14}  the convergence is still under a question: MR$_{3}$-CISD and MR$_{12}$-CISD methods are stable with respect to basis set change while the MR$_{9}$-CISD does not.

Note also, that in Ref.~\cite{Skripnikov:16b} we have found that the CVTZ basis set \cite{Dyall:07,Dyall:12} is not complete enough.
Addition of $g-$, $h-$ and $i-$ type Gaussians to the basis set changes \Eeff\ by about 0.6 GV/cm.

\subsection{Core correlation}
In \cite{Skripnikov:15a} we have analyzed correlation contributions to \Eeff\ and other parameters of ThO from the outer-core electrons of Th ($5s^25p^65d^{10}$) and O ($1s^2$).
We have concluded:
(i) contribution of correlation of the outer-core electrons is about 4 GV/cm;
(ii) the 5 a.u.\ cut-off by energy for virtual spinors is insufficient to get the correct estimation of the contribution; In the latter case the contribution decreases to 3.3 GV/cm.
Qualitatively, this was explained by the absence of important correlation functions in the considered small virtual space to describe correlation of the outer core electrons.
(iii) Contribution from correlation of the outer-core electrons can not be considered within the MR$_{3}$-CISD method because it is not size-extensive \cite{Bartlett:95}, that is the correlation energy of the method does not scale properly with the number of electrons.
(iv) size-extensive and size-consistent coupled cluster methods are stable with respect to increasing of the level of included cluster amplitudes (CCSD(T), CCSDT and CCSDT(Q) give almost the same contributions).

In \cite{FleigComment:16} the authors checked point (ii) and found that within the MR$_{3}$-CISD method the 5 a.u. cut-off for virtual spinors is indeed insufficient. Increasing cut-off to 38 Hartrees they obtained contribution of +1.2 GV/cm within this method. Taking into account that the method is not size-extensive we cannot go to a conclusion that the contribution of the outer-core electrons is now well-saturated in \cite{FleigComment:16}.

In Ref.~\cite{Skripnikov:16b} we have confirmed that the MR$_{3}$-CISD method indeed cannot be used for accurate estimation of the outer-core contribution. As we move from the MR$_{3}$-CISD to the the MR$_{3}$-CISDT method the outer-core contribution changes by more than 100\% (see \cite{Skripnikov:16b} for details).

\subsection{Subvalence and valence correlations}
In the paper \cite{FleigComment:16} the authors have extended their base model of accounting for electron correlation. Instead of the MR$_{12}$-CISD(18) model they suggested to use the MR$_{12}^{+T}$-CISD(18) model. In this model in addition to the excitations of the MR$_{12}$-CISD(18) method a new set of important type of excitations has been included in which three holes in the Th 6s, 6p and O 2s, 2p outer-core spinors are allowed. This new type of correlation effects have increased their previous base value of \Eeff\ by 2.5\% (+1.9 GV/cm) \cite{FleigComment:16}.

We would like to note that some types of correlation effects are missed or treated only partly in the new model.
For example, the four-fold excitations from the closed-shell spinors (Th 6s, 6p and O 2s, 2p) to open-shell and virtual space are not included. According to our estimates \cite{Skripnikov:15a} this type of correlation can give about 0.8 GV/cm to the value of \Eeff. Sure, it is within the stated theoretical uncertainty. 
One more note is that the three-fold excitations from 16 closed-shell spinors to open-shell and virtual ones as well as simultaneous three-fold excitations from the 16 closed-shell spinors and one-fold open-shell to virtual are not fully described by the MR$_{12}^{+T}$-CISD(18) model.
According to our estimates these kinds of excitations 
can significantly contribute to \Eeff.
Thus, it is not possible to answer whether the leading contribution from these type of excitations has been treated by the MR$_{12}^{+T}$-CISD(18).

Authors of \cite{FleigComment:16} have also considered correction on expansion of the active spinor space for their new base model MR$_{K}^{+T}$-CISD(18) with K=12. For this they took difference of the values of the considered parameters within the MR$_{31}$-CISD(18) and MR$_{12}$-CISD(18) methods. One can interpret the sum of values obtained within the MR$_{12}^{+T}$-CISD(18) method and the ``(MR$_{31}$-CISD(18) - MR$_{12}$-CISD(18))'' correction as an approximation to results of MR$_{31}^{+T}$-CISD(18) method. 
However, one should take account of the results in Table \ref{TFlResults} compiled from \cite{Fleig:14,FleigComment:16}. One can see that the contribution of the new type of excitations within the MR$_{K}^{+T}$-CISD(18) method with respect to the MR$_{K}$-CISD(18) significantly increase with expansion of the active space K from K=3 to K=12, i.e. the correction for \Eeff\  increases from -0.2 GV/cm for K=3 to +1.9 for K=31. One can suggest that the contribution to the considered properties from these kinds of excitations due to the expansion of the active space from K=12 to K=31 (and higher) will not be negligible.

One more note concerns convergence of the considered MR$_{K}$-CISD(18) and MR$_{K}^{+T}$-CISD(18) methods with expansion of space up to K=31. According to \cite{FleigComment:16} K=31 corresponds to inclusion of virtual orbitals with energies up to 0.527 a.u. in the active space. As was mentioned above the restriction of virtual spinor space for MR$_{K}$-CISD(36) method by +5 a.u. resulted in poor description of correlation of the outer-core electrons with energies in the range $-12$--$-4$ a.u. This was confirmed both in \cite{Skripnikov:15a} and in \cite{FleigComment:16}. In the present case the orbital energies of 18 considered electrons ($6s$, $6p$, $7s$, $6d$ of Th and $2s$, $2p$ of O) have the lower bound of energies of -2 a.u. Therefore, additional expansion of the active space can also be important.

\begin{table}[!h]
\centering
\caption{Results of property calculations from Refs. \cite{Fleig:14,FleigComment:16}}
\label{TFlResults}
\begin{tabular}{llrrr}
\# & Model                       & \Eeff, GV/cm         & A$_{||}$, MHz        & W$_S$, kHz           \\
\hline  
1           & vDZ/MR$_3$-CISD(18) (*)      & 80.4                 & -1263                & ---        \\
2           & vDZ/MR$_3^{+T}$-CISD(18) (*) & 80.2                 & -1251                & ---        \\
\hline  
3           & vTZ/MR$_{12}$-CISD(18)      & 75.2                 & -1339                & 106.0       \\
4           & vTZ/MR$_{12}^{+T}$-CISD(18) & 77.1                 & -1309                & 108.5       \\
\end{tabular}
(*) The virtual cutoff is set to 5 a.u.
\end{table}

\section{Conclusion}
In the present paper we have given a reply to the comment of \cite{FleigComment:16}.
Some of the questions that we addressed in \cite{Skripnikov:15a} to the methods used in \cite{Fleig:14} are partly replied in \cite{FleigComment:16}. In particular, taking into account new results (values of effective electric field, hyperfine structure constant, etc) from \cite{FleigComment:16} obtained within the MR$_{12}$-CISD method seems to be more stable to the variation of one-electron spinors 
than the results within the MR$_{3}$-CISD method analyzed in \cite{Skripnikov:15a}, though the uncertainty of HFS is still about 7\%.
The point about contribution of correlation of the outer-core electrons ($5s$, $5p$, $5s$ of Th) seems to us not solved in \cite{FleigComment:16} due to treatment within not size-extensive MR$_{3}$-CISD method with rather poor treatment of correlation effects.
Finally, the point about limited treatment of correlation effects within the MR$_{12}$-CISD(18) used in \cite{Fleig:14} is partly solved within the MR$_{12}^{+T}$-CISD(18) method (and the corresponding corrections) but the convergence of the correlations treatment is not fully justified.

\section{Acknowledgement}

The work was funded by RFBR, according to the research project No.~16-32-60013 mol\_a\_dk and President of the Russian Federation Grant No.~MK-7631.2016.2.


\end{document}